\documentclass[letterpaper, 10 pt, conference]{ieeeconf}
\IEEEoverridecommandlockouts
\usepackage{amsmath, amssymb, bm, graphicx, cite, url}
\usepackage[most]{tcolorbox}

\usepackage{cleveref}
\crefformat{equation}{(#2#1#3)}
\Crefname{algocfline}{Algorithm}{Algorithms}
\Crefname{algocf}{line}{lines}
\Crefname{AlgoLine}{Line}{Lines}
\crefname{AlgoLine}{line}{lines}

\usepackage{xcolor}

\title{\LARGE \bf
Scaling Chance-Constrained Correlated Equilibrium Computation \\
for Exclusive Resource-Assignment Games
}

\author{Jaehan Im, Ufuk Topcu, and David Fridovich-Keil
\thanks{This work was supported by the NSF under grants 2336840 and 2211548, by NASA under ULI grants 80NSSC21M0071 and 80NSSC24M0070, and by ONR under grant N00014-22-1-2703.}
\thanks{Jaehan Im, Ufuk Topcu and David Fridovich-Keil are with the Department of Aerospace Engineering, The University of Texas at Austin,
        Austin, TX 78712, USA
        {\tt\small jaehan.im@utexas.edu; utopcu@utexas.edu; dfk@utexas.edu}}%
}

\begin{document}
\bstctlcite{BSTcontrol}
\newcommand{\actSet}{\mathcal X}
\newcommand{\sys}{\text{sys}}

\newcommand{\ones}{\bm 1}
\newcommand{\reals}{{\mbox{\bf R}}}
\newcommand{\integers}{{\mbox{\bf Z}}}
\newcommand{\symm}{{\mbox{\bf S}}}  
\newcommand{\lag}{\mathcal{L}}

\newcommand{\nullspace}{{\mathcal N}}
\newcommand{\range}{{\mathcal R}}
\newcommand{\Rank}{\mathop{\bf Rank}}
\newcommand{\Tr}{\mathop{\bf Tr}}
\newcommand{\diag}{\mathop{\bf diag}}
\newcommand{\card}{\mathop{\bf card}}
\newcommand{\rank}{\mathop{\bf rank}}
\newcommand{\conv}{\mathop{\bf conv}}
\newcommand{\prox}{\bm{prox}}

\newcommand{\Expect}{\mathop{\bf E{}}}
\newcommand{\Prob}{\mathop{\bf Prob}}
\newcommand{\Co}{{\mathop {\bf Co}}} 
\newcommand{\dist}{\mathop{\bf dist{}}}
\newcommand{\argmin}{\mathop{\rm argmin}}
\newcommand{\argmax}{\mathop{\rm argmax}}
\newcommand{\epi}{\mathop{\bf epi}} 
\newcommand{\Vol}{\mathop{\bf vol}}
\newcommand{\dom}{\mathop{\bf dom}} 
\newcommand{\intr}{\mathop{\bf int}}
\newcommand{\sign}{\mathop{\bf sign}}
\newcommand{\norm}[1]{\left\lVert#1\right\rVert}
\newcommand{\mnorm}[1]{{\left\vert\kern-0.25ex\left\vert\kern-0.25ex\left\vert #1 
    \right\vert\kern-0.25ex\right\vert\kern-0.25ex\right\vert}}



\newtheorem{definition}{Definition} 
\newtheorem{theorem}{Theorem}
\newtheorem{lemma}{Lemma}
\newtheorem{corollary}{Corollary}
\newtheorem{remark}{Remark}
\newtheorem{proposition}{Proposition}
\newtheorem{assumption}{Assumption}
\newtheorem{example}{Example}

\newcommand{\cf}{{\it cf.}}
\newcommand{\eg}{{\it e.g.}}
\newcommand{\ie}{{\it i.e.}}
\newcommand{\etc}{{\it etc.}}

\newcommand{\ba}[2][]{\todo[color=orange!40,size=\footnotesize,#1]{[BA] #2}}

\newcommand{\fix}[1]{{#1}}

\newcommand{\bigO}{\mathcal{O}}

\newcommand{\intSet}{\mathbb{Z}}
\newcommand{\realSet}{\mathbb{R}}
\newcommand{\natSet}{\mathbb{N}}
\newcommand{\zeroSet}{\bm{0}}
\newcommand{\state}{\bm{x}}
\newcommand{\cmdh}{\bar{\bm{u}}}
\newcommand{\cmda}{\mathring{\bm{u}}}
\newcommand{\cmd}{\bm{u}}
\newcommand{\costh}{J_h}
\newcommand{\costa}{J_a}
\newcommand{\observ}{\bm{z}}

\maketitle
\thispagestyle{empty}
\pagestyle{empty}

\begin{abstract}
The chance-constrained correlated equilibrium concept provides a robust coordination technique for self-interested agents whose costs are uncertain to a central coordinator, but the number of joint actions considered in its computation grows exponentially with the number of agents. 
\fix{We develop a reduced-rank formulation based on convex combinations of chance-constrained pure Nash equilibria to improve scalability. To further reduce computation, we} propose an \emph{exact-one resource-assignment restriction} for games in which agents compete for shared resources and simultaneous use of a resource is undesirable or unsafe.
The restriction allows each resource to be assigned to exactly one agent, reducing the number of joint actions considered from $\mathcal{O}(2^{nr})$ to $\mathcal{O}(n^r)$ for $n$ agents and $r$ resources. 
The resulting problem is solved over the restricted joint action set while retaining all incentive constraints associated with unilateral deviations. 
We show that the feasible set of the restricted problem is exactly the convex hull of the chance-constrained pure Nash equilibria contained in the exact-one assignment set, yielding a necessary and sufficient condition for nonemptiness. 
We further derive a sufficient condition under which all chance-constrained pure Nash equilibria of the unrestricted game are retained by the restriction. 
Numerical experiments in a vertiport departure-corridor coordination scenario demonstrate a $99.81\%$ reduction in median computation time relative to the unrestricted formulation while achieving equivalent coordination performance.
\end{abstract}

\section{Introduction}

Large-scale multi-agent systems often consist of self-interested agents whose decisions collectively determine system-level outcomes. 
The correlated equilibrium (CE) concept provides a coordination mechanism by allowing a coordinator to recommend actions from which agents have no incentive to unilaterally deviate \cite{aumann,aumann_2}. 
When agents' costs are uncertain because of modeling errors, operational noise, or private information, the chance-constrained correlated equilibrium (CCCE) concept extends this framework by requiring incentive compatibility to hold at a prescribed confidence level \cite{j_ccce}.

The CE and CCCE formulations, however, face a scalability limitation because they optimize a probability distribution over the joint action space. With $n$ agents and at most $m$ actions per agent, the number of joint action probability variables grows as $\mathcal{O}(m^n)$. Consequently, explicitly constructing and solving the CE or CCCE formulation becomes impractical as $n$ grows \cite{rrce}.

The reduced-rank correlated equilibrium (RRCE) concept mitigates this limitation using Nash-equilibrium representations \cite{rrce}. 
A mixed Nash equilibrium admits a product-form representation with $\mathcal{O}(nm)$ variables, and convex combinations of finitely many such equilibria provide a lower-dimensional representation of a subset of the CE set. 
\fix{We develop a reduced-rank formulation for CCCE by constructing} the reduced representation from chance-constrained pure Nash equilibria (CCPNEs), each of which individually satisfies the chance-constrained incentive conditions, and \fix{optimizing} over their convex hull. 
However, although \fix{this formulation reduces} the optimization dimension, CCPNE existence is not guaranteed in general, and exhaustively identifying all CCPNEs may still require examining $\mathcal{O}(m^n)$ joint action profiles.

\begin{figure}[t!]
    \centering
    \includegraphics[width=0.95\linewidth, trim = 0 0.8cm 0 0]{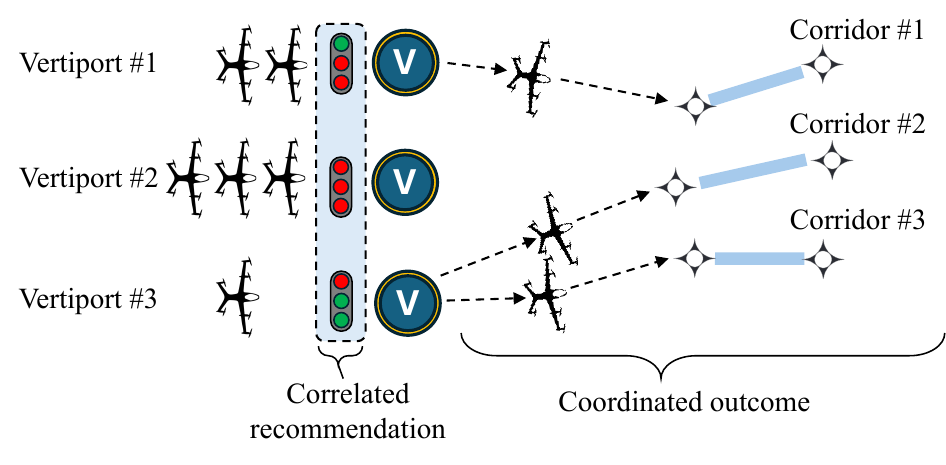}
    \caption{Illustration of the departure-corridor coordination scenario. Independently operated vertiports act as self-interested agents that determine whether to occupy or yield each shared departure corridor. A central coordinator broadcasts correlated recommendations that suggest corridor usage while preserving the vertiports' autonomy.}
    \label{fig:concept}
\end{figure}

We propose an \emph{exact-one resource-assignment restriction} for exclusive resource-assignment games, in which agents compete for shared resources and simultaneous use of a resource by multiple agents is undesirable.
Such structures often arise in intersection management, aircraft terminal-area coordination, and vertiport resource allocation \cite{levin2017conflict,xue2014optimal,espejo2023heuristic}. 
With $r$ binary resource-occupancy decisions per agent, the unrestricted joint action space contains $\mathcal{O}(2^{nr})$ profiles. 
The exact-one restriction assigns each resource to exactly one of the $n$ agents, reducing the number of profiles considered by the coordinator to $\mathcal{O}(n^r)$, which is polynomial in $n$ for fixed $r$. 
The resulting CCCE problem can be solved directly without enumerating CCPNEs while retaining all incentive constraints associated with unilateral deviations. 
\Cref{fig:concept} illustrates the restriction in a vertiport departure-corridor coordination setting.

The contributions of this paper are threefold. 
First, we \fix{develop a reduced-rank CCCE formulation based on convex combinations of CCPNEs and} introduce an exact-one resource-assignment restriction that reduces CCCE computation from $\mathcal{O}(2^{nr})$ to $\mathcal{O}(n^r)$ joint actions while retaining all unilateral-deviation incentive constraints.
Second, we characterize the feasibility of the restricted formulation and derive a sufficient condition for retaining all CCPNEs in the unrestricted joint action space. Third, we compare the proposed restriction with unrestricted and reduced-rank formulations in an advanced air mobility coordination scenario.



\section{Related work}

\subsection{Correlated equilibrium computation and scalability}
The \fix{CE} concept has been widely studied as a mechanism for coordinating noncooperative agents while preserving individual autonomy \cite{ce_app_2}. Uncertainty in payoff or cost models has been addressed through Bayes correlated equilibrium \cite{ce1}, robust and distributionally robust game formulations \cite{ce2,ce4,ce8}, and bounded rationality models such as quantal response equilibrium \cite{qre1,qre2}. 
The \fix{CCCE} concept instead imposes a probabilistic threshold on incentive compatibility under stochastic cost perturbations \cite{j_ccce}.

For games represented in normal form, CE can be computed through linear programming. However, the formulation contains one probability variable per joint action, so the number of variables grows exponentially with the number of agents. Learning dynamics \cite{tract_1,tract_2} and approximation methods \cite{tract_3} can avoid directly solving the original CE program, although they are not generally designed for targeted coordinator optimization over the CE set.

\subsection{Reduced representations and scalable computation}

The \fix{RRCE} concept restricts the joint action distribution to the convex hull of a finite collection of mixed Nash equilibria \cite{rrce}. 
\fix{This approach reduces} the optimization dimension, but the Nash equilibria forming the reduced representation must first be identified, which may still require searching a large joint action space.

Other scalable approaches exploit compact game representations \cite{papadimitriou2008computing}, generate correlated-strategy columns through pricing procedures \cite{zhang2026optimal}, or iteratively add constraints through separation procedures \cite{dehghanian2024identifying}. In contrast, the present approach constructs a restricted joint action set a priori from the combinatorial structure of the resource-assignment problem, without iterative column or constraint generation. 
This restriction yields a directly specified reduced formulation and permits explicit characterization of its feasibility.

\section{Preliminaries}
\label{sec:preliminaries}


\subsection{Correlated equilibrium and coordination}

Consider a finite game with agent set $\mathcal N:=\{1,\ldots,n\}$. Agent $i\in\mathcal N$ selects an action $x_i$ from a finite set $\mathcal X_i$, with $|\mathcal X_i|=m$, and the joint action space is $\mathcal X:=\prod_{i\in\mathcal N}\mathcal X_i$.
Let $J_i:\mathcal X\rightarrow\mathbb R$ denote the cost of agent $i$. For a recommended action $a_i\in\mathcal X_i$ and a deviating action $a_i'\in\mathcal X_i\setminus\{a_i\}$, define the deviation margin
\begin{equation}
    \textstyle \Delta J_i(a_i,a_i',x_{\neg i}):=J_i(a_i,x_{\neg i})-J_i(a_i',x_{\neg i}),
    \label{eq:ce_deviation_margin}
\end{equation}
\fix{where $x_{\neg i}$ denotes the joint actions of all agents other than agent $i$.}
A positive margin means that the deviation decreases the agent's cost.

\begin{definition}[Correlated equilibrium \cite{aumann_2}]
Let $z\in\Delta(\mathcal X)$ denote a probability distribution over joint actions. A \fix{CE} satisfies
\begin{equation}
    \textstyle \sum_{x_{\neg i}}z(a_i,x_{\neg i})\Delta J_i(a_i,a_i',x_{\neg i})\leq 0
    \label{eq:ce_constraint}
\end{equation}
for every $i$, $a_i$, and $a_i'$.
\end{definition}

Given a coordinator objective $J_{\rm sys}:\Delta(\mathcal X)\rightarrow\mathbb R$, CE-based coordination solves
\begin{equation}
    \textstyle \min_{z\in\Delta(\mathcal X)} J_{\rm sys}(z)\quad\text{s.t.}\quad z\text{ satisfies~\Cref{eq:ce_constraint}},
    \label{eq:ce_coordination_problem}
\end{equation}
which involves $m^n$ probability variables \cite{rrce}.

\subsection{Chance-constrained correlated equilibrium}

Suppose the coordinator has access to nominal costs $\bar J_i:\mathcal X\rightarrow\mathbb R$ and models uncertainty in an agent's deviation margin as
\begin{equation}
    \textstyle \Delta J_i(a_i,a_i',x_{\neg i})=\Delta\bar J_i(a_i,a_i',x_{\neg i})+\eta_i,
    \label{eq:uncertain_deviation_margin}
\end{equation}
where $\eta_i\sim\mathcal N(0,\sigma_i^2)$, $\sigma_i$ denotes the deviation-margin uncertainty for agent $i$, and $\Delta\bar J_i(a_i,a_i',x_{\neg i})$ is defined analogously to \Cref{eq:ce_deviation_margin}. For a fixed confidence level $\alpha\geq 1/2$, let
\begin{equation}
    \textstyle \kappa_i:=\Phi^{-1}(\alpha)\sigma_i\geq 0,
    \label{eq:kappa_definition}
\end{equation}
where $\Phi$ is the standard normal cumulative distribution function.

\begin{definition}[Chance-constrained CE \cite{j_ccce}]
Under the uncertainty model in~\Cref{eq:uncertain_deviation_margin}, a probability distribution $z\in\Delta(\mathcal X)$ is a \fix{CCCE} if
\begin{equation}
    \textstyle \sum_{x_{\neg i}}z(a_i,x_{\neg i})\left(\Delta\bar J_i(a_i,a_i',x_{\neg i})+\kappa_i\right)\leq 0
    \label{eq:ccce_constraint}
\end{equation}
for every $i$, $a_i$, and $a_i'$.
\end{definition}

Define the CCCE constraint matrix $A_{\mathcal X}$ by indexing each row by $q=(i,a_i,a_i')$ and each column by $x\in\mathcal X$:
\begin{equation}
    \textstyle [A_{\mathcal X}]_{q,x}:=\mathbf 1\{x_i=a_i\}\left(\Delta\bar J_i(a_i,a_i',x_{\neg i})+\kappa_i\right),
    \label{eq:full_ccce_matrix}
\end{equation}
where $\mathbf 1\{\cdot\}$ denotes the indicator function. The CCCE feasible set and coordination problem are
\begin{align}
    \textstyle \mathcal Z_{\mathcal X} &\textstyle :=\left\{z\in\Delta(\mathcal X):A_{\mathcal X}z\leq 0\right\},
    \label{eq:full_ccce_set}
    \\
    \textstyle J_{\mathcal X}^\star &\textstyle :=\min_{z\in\mathcal Z_{\mathcal X}}J_{\rm sys}(z).
    \label{eq:full_ccce_problem}
\end{align}

\section{\fix{Reduced-rank CCCE formulation}}
The reduced-rank approach improves scalability by computing Nash equilibria and optimizing over their convex hull \cite{rrce}.
\fix{We extend this approach to CCCE by optimizing over convex combinations of CCPNEs.}

A joint action $x\in\mathcal X$ corresponds to the degenerate
distribution $e_x$, which places unit probability on $x$. Substituting $z=e_x$ into \Cref{eq:ccce_constraint} reduces the CCCE condition to
\begin{equation}
    \Delta\bar J_i(x_i,a_i',x_{\neg i})+\kappa_i\leq 0,
    \label{eq:ccpne_definition}
\end{equation}
for every agent $i$ and every
$a_i'\in\mathcal X_i\setminus\{x_i\}$.
A joint action satisfying this condition is a \fix{CCPNE}.

Let $\mathcal P_{\mathcal X}$ denote the set of all CCPNEs in $\mathcal X$.
Every $x\in\mathcal P_{\mathcal X}$ induces a CCCE, and by convexity of $\mathcal Z_{\mathcal X}$\fix{, which follows from its linear defining constraints},
\begin{equation}
    \operatorname{conv}\{e_x:x\in\mathcal P_{\mathcal X}\}
    \subseteq
    \mathcal Z_{\mathcal X},
    \label{eq:pne_hull_in_full_ccce}
\end{equation}
where $\operatorname{conv}$ denotes the convex hull.
Given $d$ precomputed CCPNEs
$\mathcal P_d:=\{x^{(1)},\ldots,x^{(d)}\}\subseteq\mathcal P_{\mathcal X}$,
the reduced-rank CCCE approach solves
\begin{equation}
    \textstyle
    \min_{\lambda\in\Delta_d}
    J_{\rm sys}\!\left(
        \sum_{k=1}^{d}\lambda_k e_{x^{(k)}}
    \right).
    \label{eq:rr_ccce_problem}
\end{equation}
This problem contains $d$ mixture variables and requires no CCCE constraints because each vertex, and therefore their convex hull, consists of certified CCCEs. However, the reduction retains the cost of finding and verifying the CCPNEs, whose existence and number are not guaranteed in a general game.

\section{Exact-one resource-assignment restriction\\ for an exclusive resource-assignment game}
\label{sec:restriction}

\subsection{Exclusive resource-assignment game}

Let $\mathcal V:=\{1,\ldots,r\}$ be a set of resources, where $r\geq 1$.
Each agent selects a binary resource-occupancy vector
\begin{equation}
    \textstyle x_i=(x_{i1},\ldots,x_{ir})\in\mathcal X_i:=\{0,1\}^r,
    \label{eq:binary_resource_action}
\end{equation}
where $x_{iv}=1$ indicates that agent $i$ occupies resource $v$.
Thus, the number of actions available to each agent is $m=2^r$, and the unrestricted joint action set contains $|\mathcal X|=(2^r)^n=2^{nr}$ profiles. 
For each resource $v$, define its occupancy count as $N_v(x):=\sum_{i\in\mathcal N}x_{iv}$.
Agent $i$ has nominal cost
\begin{equation}
    \textstyle \bar J_i(x)
    =
    c\sum_{v\in\mathcal V}x_{iv}\bigl(N_v(x)-1\bigr)
    +
    \rho_i\sum_{v\in\mathcal V}(1-x_{iv}),
    \label{eq:resource_assignment_cost}
\end{equation}
where $\rho_i\in\realSet_{>0}$ is agent $i$'s resource-yielding cost and
$c\in\realSet_{>0}$ is a common conflict penalty.

\subsection{Exact-one resource-assignment restriction}

We restrict the coordinator's recommendations to the exact-one assignment set
\begin{equation}
    \textstyle \mathcal F
    :=
    \left\{
        x\in\mathcal X:
        N_v(x)=1,\quad \forall v\in\mathcal V
    \right\}.
    \label{eq:exact_one_set}
\end{equation}
Each resource is assigned independently to one of $n$ agents, so
$|\mathcal F|=n^r$.

Let $A_{\mathcal F}$ be the submatrix of $A_{\mathcal X}$ containing the
columns indexed by $x\in\mathcal F$. The restricted CCCE feasible set is
\begin{equation}
    \textstyle \mathcal Z_{\mathcal F}
    :=
    \left\{
        z\in\Delta(\mathcal F):
        A_{\mathcal F}z\leq 0
    \right\},
    \label{eq:restricted_ccce_set}
\end{equation}
and the proposed coordination problem is
\begin{equation}
    \textstyle J_{\mathcal F}^\star
    :=
    \min_{z\in\mathcal Z_{\mathcal F}}J_{\rm sys}(z).
    \label{eq:restricted_ccce_problem}
\end{equation}
The number of probability variables is reduced from
$\mathcal O(2^{nr})$ to $\mathcal O(n^r)$. Hence, the dependence on $n$ is
polynomial for fixed $r$, although the formulation remains exponential in $r$. Unlike the reduced-rank approach \Cref{eq:rr_ccce_problem}, the restricted formulation \Cref{eq:restricted_ccce_problem} requires no prior CCPNE enumeration.
\medskip

\begin{example}[Illustration of the assignment restriction]
Consider a simple intersection with two vehicles approaching on conflicting
paths, where each vehicle chooses either to proceed ($G$) or stop ($S$).
For $n=2$ and $r=1$, the cost model in~\Cref{eq:resource_assignment_cost}
gives
\small
\begin{equation}
\textstyle
\label{eq:intersection_game}
\begin{array}{c|cc}
 & G & S \\ \hline
G & (c,c) & (0,\rho_2) \\
S & (\rho_1,0) & (\rho_1,\rho_2)
\end{array}.
\end{equation}
\normalsize
The joint action $GG$ produces a conflict, whereas $SS$ leaves the shared
resource unused. The exact-one restriction therefore retains only
$\mathcal F=\{GS,SG\}$, corresponding to conflict-free and resource-utilizing
recommendations.
\end{example}


\section{Feasibility and equilibrium properties}
\label{sec:theory}
A natural question arising from the joint action restriction is whether the restricted \fix{CCCE} problem in \Cref{eq:restricted_ccce_problem} remains feasible and \fix{which equilibria} are lost by restricting the joint action space.
We characterize the feasibility of the restricted CCCE and establish conditions under which the restriction retains all \fix{CCPNEs} in the unrestricted joint action space.

\subsection{CCPNE characterization within the exact-one set}

For $x\in\mathcal F$, define the set of resources assigned to agent $i$ as $S_i(x):=\{v\in\mathcal V:x_{iv}=1\}$.
We first characterize when a joint action in the exact-one set is a CCPNE.

\begin{lemma}[CCPNE characterization in $\mathcal F$]
\label{lem:restricted_structure}
For any $x\in\mathcal F$ and $x_i'\neq x_i$, the deviation margin
$\Delta\bar J_i(x_i,x_i',x_{\neg i})$ is independent of $x_{\neg i}$.
Moreover, $x$ is a CCPNE if and only if, for every
$i\in\mathcal N$,
\begin{equation}
\begin{aligned}
    \textstyle S_i(x)\neq\emptyset
    &\textstyle \Longrightarrow \kappa_i\leq\rho_i,\\
    \textstyle S_i(x)\neq\mathcal V
    &\textstyle \Longrightarrow \kappa_i\leq c-\rho_i.
\end{aligned}
\label{eq:profile_ccpne_conditions}
\end{equation}

\begin{proof}
Since $x\in\mathcal F$, each resource is occupied by exactly one agent.
For a deviation $x_i'\neq x_i$, let $D_i^-:=\{v:x_{iv}=1,\ x_{iv}'=0\}$, and $\textstyle D_i^+:=\{v:x_{iv}=0,\ x_{iv}'=1\}$.
Giving up a resource changes agent $i$'s cost from $0$ to $\rho_i$,
whereas taking a resource already assigned to another agent changes its cost from
$\rho_i$ to $c$. Hence,
\begin{equation}
\textstyle \Delta\bar J_i
=
-\rho_i|D_i^-|-(c-\rho_i)|D_i^+|.
\label{eq:restricted_deviation_margin}
\end{equation}
Thus, the deviation margin is independent of $x_{\neg i}$.

If $S_i(x)\neq\emptyset$, giving up a single resource yields the CCPNE condition in~\Cref{eq:ccpne_definition} $-\rho_i+\kappa_i\leq0$; similarly, if $S_i(x)\neq\mathcal V$, taking a single resource yields $-(c-\rho_i)+\kappa_i\leq0$.
Thus, \Cref{eq:profile_ccpne_conditions} is necessary.
Conversely, under \Cref{eq:profile_ccpne_conditions}, any nontrivial
deviation ($|D_i^-|+|D_i^+| \geq 1$) satisfies
\begin{equation}
\textstyle \Delta\bar J_i+\kappa_i
\leq
-\bigl(|D_i^-|+|D_i^+|-1\bigr)\kappa_i
\leq0,
\end{equation}
so $x$ is a CCPNE.
\end{proof}
\end{lemma}

\subsection{Restricted feasibility and CCPNE existence}
\label{sec:restricted_feasibility}

We next derive a necessary and sufficient condition for the restricted CCCE
feasible set in \Cref{eq:restricted_ccce_set} to be nonempty.
Define the CCPNEs contained in the exact-one assignment set as
$\mathcal P_{\mathcal F}:=\mathcal P_{\mathcal X}\cap\mathcal F$.

\begin{theorem}[Restricted feasibility--CCPNE equivalence]
\label{thm:restricted_geometry}
For the cost structure in~\Cref{eq:resource_assignment_cost},
\begin{equation}
\textstyle \mathcal Z_{\mathcal F}
=
\operatorname{conv}
\{e_x:x\in\mathcal P_{\mathcal F}\},
\label{eq:restricted_set_equals_pne_hull}
\end{equation}
where $e_x$ is the degenerate distribution placing unit probability on $x$.
Consequently,
\begin{equation}
\textstyle \mathcal Z_{\mathcal F}\neq\emptyset
\quad\Longleftrightarrow\quad
\mathcal P_{\mathcal F}\neq\emptyset.
\label{eq:restricted_feasible_iff_pne}
\end{equation}

\begin{proof}
Let $z\in\mathcal Z_{\mathcal F}$ and consider any profile
$x\in\mathcal F$ satisfying $z(x)>0$. For every agent $i$, define
\begin{equation}
\textstyle p_i(x_i;z)
:=
\sum_{\tilde x_{\neg i}:
(x_i,\tilde x_{\neg i})\in\mathcal F}
z(x_i,\tilde x_{\neg i})
\geq z(x)>0,
\label{eq:positive_recommendation_probability}
\end{equation}
\fix{where $\tilde x_{\neg i}$ indexes the other agents'
joint actions compatible with $x_i$ in $\mathcal F$.}
By~\Cref{lem:restricted_structure}, for any fixed deviation $x_i'\neq x_i$,
the deviation margin is independent of $\tilde x_{\neg i}$ over these
profiles. Hence, the corresponding CCCE constraint in
\Cref{eq:ccce_constraint} can be written as
\begin{equation}
\textstyle p_i(x_i;z)
\left(
\Delta\bar J_i(x_i,x_i',x_{\neg i})+\kappa_i
\right)
\leq0.
\label{eq:factorized_restricted_constraint}
\end{equation}
\fix{Since $p_i(x_i;z)>0$, \Cref{eq:factorized_restricted_constraint} implies the CCPNE condition in~\Cref{eq:ccpne_definition}.}
Hence, $x$ is a CCPNE, and since $x\in\mathcal F$,
$x\in\mathcal P_{\mathcal F}$.
Since $x$ was arbitrary in the positive support of $z$,
\begin{equation}
\textstyle z = \sum_{x\in\mathcal P_{\mathcal F}}z(x)e_x
\in
\operatorname{conv}\{e_x:x\in\mathcal P_{\mathcal F}\}.
\end{equation}

Conversely, every $x\in\mathcal P_{\mathcal F}$ induces a feasible
degenerate CCCE $e_x$. Because the CCCE constraints are linear, every
convex combination of these distributions also belongs to
$\mathcal Z_{\mathcal F}$. This proves
\Cref{eq:restricted_set_equals_pne_hull}, from which
\Cref{eq:restricted_feasible_iff_pne} follows.
\end{proof}
\end{theorem}

We have shown that the restricted CCCE feasibility is equivalent
to CCPNE existence in $\mathcal F$ by \Cref{thm:restricted_geometry}. We now use~\Cref{lem:restricted_structure} to derive an existence condition.

\begin{theorem}[Restricted CCPNE existence condition]
\label{thm:canonical_pne_existence}
$\mathcal P_{\mathcal F}$ is nonempty if and only if there exists
$h\in\mathcal N$ such that
\begin{equation}
\textstyle \kappa_h\leq\rho_h,
\qquad
\kappa_j\leq c-\rho_j,
\quad\forall j\neq h.
\label{eq:explicit_feasibility_condition}
\end{equation}

\begin{proof}
Suppose $\mathcal P_{\mathcal F}\neq\emptyset$ and let
$x\in\mathcal P_{\mathcal F}$. Select any resource $v\in\mathcal V$ and
let $h$ denote its unique owner. Then $S_h(x)\neq\emptyset$, and
\Cref{lem:restricted_structure} implies $\kappa_h\leq\rho_h.$
For every $j\neq h$, resource $v$ is not assigned to $j$, so
$S_j(x)\neq\mathcal V$. Hence, $\textstyle \kappa_j\leq c-\rho_j$, for all $j\neq h$.
Thus, \Cref{eq:explicit_feasibility_condition} is necessary.

Conversely, suppose there exists $h\in\mathcal N$ satisfying
\Cref{eq:explicit_feasibility_condition}. Consider the profile $x$ that assigns
every resource to agent $h$. Then $S_h(x)=\mathcal V$ and
$S_j(x)=\emptyset$ for every $j\neq h$. Thus,
\Cref{eq:profile_ccpne_conditions} is satisfied for every agent, and
$x\in\mathcal P_{\mathcal F}$, so
$\mathcal P_{\mathcal F}\neq\emptyset$.
\end{proof}
\end{theorem}



By \Cref{thm:restricted_geometry,thm:canonical_pne_existence}, the restricted CCCE problem is feasible if and only if \Cref{eq:explicit_feasibility_condition} holds for some $h\in\mathcal N$.

\subsection{Recovery of CCPNEs in $\mathcal X$}
\label{sec:pne_recovery}

The preceding results characterize the feasibility of the restricted problem \Cref{eq:restricted_ccce_problem}. We now investigate when restricting the joint action space to $\mathcal F$ preserves all CCPNEs in $\mathcal X$.

\begin{proposition}[CCPNE completeness of the restriction]
\label{prop:pne_completeness}
Suppose $\rho_i>0$, $\kappa_i\geq0$, and the following \emph{CCPNE recovery condition} holds:
\begin{equation}
\textstyle c-\rho_i+\kappa_i>0,
\qquad \forall i\in\mathcal N.
\label{eq:pne_recovery_condition}
\end{equation}
Then every CCPNE in $\mathcal X$ belongs to $\mathcal F$. Consequently,
\begin{equation}
\textstyle \mathcal P_{\mathcal X}
=
\mathcal P_{\mathcal F}.
\label{eq:full_restricted_pne_equality}
\end{equation}

\begin{proof}
We show that, under~\Cref{eq:pne_recovery_condition}, a CCPNE cannot
contain any resource with either zero or multiple occupancy. Since
$N_v(x)$ is integer-valued, this leaves only $N_v(x)=1$ for every
resource, which implies $x\in\mathcal F$.

Let $x\in\mathcal P_{\mathcal X}$ and consider any resource $v$.
If $N_v(x)=0$, an agent can occupy $v$ alone. By the CCPNE condition
in~\Cref{eq:ccpne_definition} \fix{and the cost function in~\Cref{eq:resource_assignment_cost}}, this deviation gives
$\textstyle \Delta\bar J_i+\kappa_i
=
\rho_i+\kappa_i>0$,
which is trivially true since $\rho_i>0$, $\kappa_i\geq0$. 
This contradicts the CCPNE condition \Cref{eq:ccpne_definition}. Thus, $N_v(x)=0$ cannot occur.

Now suppose $N_v(x)=k\geq2$. Any agent $i$ occupying $v$ can give up
the resource, for which
\begin{equation}
    \textstyle \Delta\bar J_i+\kappa_i
    =
    c(k-1)-\rho_i+\kappa_i
    \geq
    c-\rho_i+\kappa_i.
\end{equation}
Under~\Cref{eq:pne_recovery_condition}, this quantity is positive, so
multiple occupancy cannot occur in a CCPNE either.

Therefore, every CCPNE satisfies $N_v(x)=1$ for every $v\in\mathcal V$,
and hence belongs to $\mathcal F$.
\end{proof}
\end{proposition}

A simple sufficient condition for~\Cref{eq:pne_recovery_condition} is
\begin{equation}
    \textstyle c>\max_{i\in\mathcal N}\rho_i.
    \label{eq:high_conflict_condition}
\end{equation}
This regime is natural in exclusive resource-assignment problems where conflicting use of a resource is more costly than yielding it (\eg, intersection coordination).
Thus, in such regimes, the exact-one restriction provides an efficient simplification while preserving all CCPNEs in $\mathcal X$.
\fix{Note that preserving all CCPNEs does not guarantee optimal performance over the full CCCE set.}

When $\mathcal P_{\mathcal X}=\mathcal P_{\mathcal F}$,
\Cref{thm:restricted_geometry} further gives
\begin{equation}
    \textstyle \mathcal Z_{\mathcal F}
    =
    \operatorname{conv}
    \{e_x:x\in\mathcal P_{\mathcal X}\}.
    \label{eq:full_pne_hull_recovery}
\end{equation}
Hence, the restricted CCCE exactly recovers the convex hull of all CCPNEs in $\mathcal X$. 
However, this does not imply $\mathcal Z_{\mathcal F}=\mathcal Z_{\mathcal X}$, since the unrestricted CCCE set may contain distributions that are not convex combinations of CCPNEs.

\section{Computational formulations}
\label{sec:computational_formulations}

The theoretical results connect the joint action restriction in \Cref{sec:restriction} with the reduced-rank formulation in \Cref{eq:rr_ccce_problem}. We therefore consider four computational formulations: \texttt{Full-CCCE}, \texttt{Restricted-CCCE}, \texttt{RR}$_{\mathcal X}$, and \texttt{RR}$_{\mathcal F}$.
These four formulations are compared numerically in~\Cref{sec:experiments}.

The \texttt{Full-CCCE} and \texttt{Restricted-CCCE} formulations solve \Cref{eq:full_ccce_problem} and \Cref{eq:restricted_ccce_problem}, respectively. The unrestricted formulation uses $2^{nr}$ probability variables, whereas the restricted formulation uses $n^r$ variables. Neither formulation requires prior CCPNE identification.

The unrestricted reduced-rank formulation, \texttt{RR}$_{\mathcal X}$, is
\begin{equation}
\textstyle
J_{\mathrm{RR}_{\mathcal X}}^{\star}
:=
\min_{\lambda\in\Delta(\mathcal P_{\mathcal X})}
J_{\rm sys}\left(
\sum_{x\in\mathcal P_{\mathcal X}}
\lambda(x)e_x
\right).
\label{eq:rr_x_problem}
\end{equation}
Constructing \texttt{RR}$_{\mathcal X}$ by exhaustive search requires checking $2^{nr}$ candidate profiles.
The restricted-search reduced-rank formulation, denoted \texttt{RR}$_{\mathcal F}$, is
\begin{equation}
\textstyle
J_{\mathrm{RR}_{\mathcal F}}^{\star}
:=
\min_{\lambda\in\Delta(\mathcal P_{\mathcal F})}
J_{\rm sys}\left(
\sum_{x\in\mathcal P_{\mathcal F}}
\lambda(x)e_x
\right).
\label{eq:rr_f_problem}
\end{equation}
Constructing \texttt{RR}$_{\mathcal F}$ \fix{uses \Cref{lem:restricted_structure} to directly generate the CCPNEs in $\mathcal F$}.

By~\Cref{thm:restricted_geometry}, exhaustive \texttt{RR}$_{\mathcal F}$ and \texttt{Restricted-CCCE} have the same feasible set, and therefore
$J_{\mathcal F}^{\star}
=
J_{\mathrm{RR}_{\mathcal F}}^{\star}$.
However, they have different computational processes. \texttt{Restricted-CCCE} directly solves the CCCE problem with $n^r$ probability variables and the corresponding incentive constraints, while avoiding CCPNE screening. 
In contrast, \texttt{RR}$_{\mathcal F}$ first \fix{generates CCPNEs} and then solves a smaller $|\mathcal P_{\mathcal F}|$-variable problem without CCCE constraints.

Under the CCPNE recovery condition in~\Cref{eq:pne_recovery_condition},
$J_{\mathcal F}^{\star}
=
J_{\mathrm{RR}_{\mathcal F}}^{\star}
=
J_{\mathrm{RR}_{\mathcal X}}^{\star}$.
Thus, the joint action restriction reduces both the CCCE variable space and the CCPNE search space.

\section{Numerical experiments}
\label{sec:experiments}

\subsection{Departure-corridor coordination scenario}
\label{sec:vertiport_scenario}

\fix{Motivated by vertiport traffic coordination
\cite{xue2021coordination}, we} consider $n$ independently
operated vertiports that compete for access to $r$ shared
departure corridors over a coordination epoch, as illustrated
in~\Cref{fig:concept}.\footnote{\fix{The optimization problems in
\Cref{sec:computational_formulations} were formulated in Julia and solved with Ipopt 3.14 \cite{wachter2006implementation}.
Experiments were performed on a Mac mini with an Apple
M4 Pro chip and 24 GB of RAM.
Code is available at \url{https://github.com/CLeARoboticsLab/CCCE}.}}
Each vertiport is controlled by a self-interested operator that decides whether to dispatch an aircraft through each corridor. Specifically, $x_{iv}=1$ indicates that vertiport $i$ attempts to occupy corridor $v$, producing the binary action in~\Cref{eq:binary_resource_action}. 

Each vertiport's nominal cost follows~\Cref{eq:resource_assignment_cost}. Simultaneous use of a corridor by multiple vertiports incurs the common conflict penalty $c$, while not occupying corridor $v$ incurs the vertiport-specific delay cost $\rho_i$. The chance margin $\kappa_i$ represents uncertainty in the coordinator's estimate of vertiport $i$'s deviation incentives. The exact-one set $\mathcal F$ therefore represents conflict-free and resource-utilizing coordination in which every corridor is assigned to exactly one vertiport.

For any probability distribution $z$ over a joint action set $\mathcal A\subseteq\mathcal X$, define the expected nominal cost of agent $i$ as $C_i(z):=\sum_{x\in\mathcal A}z(x)\bar J_i(x)$,
and let $C_{\max}(z):=\max_{i\in\mathcal N}C_i(z)$. Following the fairness-threshold criterion introduced in~\cite{rrce,fairnessGuide}, the coordinator minimizes
\begin{equation}
    \small
    \textstyle J_{\rm sys}^{\Delta}(z):=-n\Delta+\sum_{i\in\mathcal N}\max\left\{C_i(z)+\Delta,C_{\max}(z)\right\},
    \label{eq:fairness_threshold_objective}
\end{equation}
\normalsize
\noindent
where $\Delta\geq0$ specifies the tolerated cost disparity. When all agent costs lie within $\Delta$ of the maximum cost,~\Cref{eq:fairness_threshold_objective} reduces to the additive cost $\sum_i C_i(z)$; larger disparities increasingly emphasize the highest agent cost. 

\begin{remark}
\label{rem:expected_allocation}
\fix{Since~\Cref{eq:fairness_threshold_objective} depends only on expected costs, the restricted problem admits an $\mathcal O(n)$-variable formulation, denoted \texttt{Compact}, when $\kappa_i\leq c-\rho_i$ for all $i$. 
It uses expected counts $y_i\geq0$, with $\sum_i y_i=r$, $y_i=0$ if $\rho_i<\kappa_i$, and $C_i=\rho_i(r-y_i)$. 
The optimal counts $y^\star$ may not uniquely determine a CCCE distribution $z$.
This reduction does not generally apply when yielding costs vary by resources or when the coordinator objective depends on quantities beyond the agents' expected costs.}
\end{remark}

\subsection{Experimental settings and metrics}
\label{sec:experimental_settings}

We fix $r=3$, giving $m=2^r=8$ actions per agent, and vary the number of vertiports over $n\in\{3,\ldots,8\}$. The common conflict penalty is fixed at $c=10$. 
For each $n$, we generate $100$ Monte Carlo instances by
independently sampling each vertiport's delay cost
\fix{uniformly from $[1,6]$}.
All agents have $\sigma_i=2$ with confidence level $\alpha=0.95$, yielding $\kappa_i\approx3.29$. The fairness threshold in~\Cref{eq:fairness_threshold_objective} is fixed at $\Delta=3$. 
\fix{We excluded 20 of 600 instances (3.3\%) without a CCPNE in $\mathcal F$ from all statistics, without resampling. By \Cref{thm:canonical_pne_existence}, the exclusion probability is $((\kappa-1)/5)^n$, giving 17.6 expected exclusions.}

We evaluate computational performance using \fix{total computation time} and the candidate-space sizes $|\mathcal X|=2^{nr}$ and $|\mathcal F|=n^r$.
\fix{Joint-action sets are generated anew for each instance, and total times include set and constraint-matrix construction and solver execution. To isolate the effect of the exact-one restriction, \texttt{RR}$_{\mathcal X}$ and \texttt{RR}$_{\mathcal F}$ use the same exhaustive CCPNE search procedure over $\mathcal X$ and $\mathcal F$, respectively.}\footnote{\fix{At $n=8$, constructing CCPNEs using Lemma~1 reduced median identification time from 2.35~ms to 0.028~ms (${\sim}85\times$) and its share of total runtime from 48\% to 0.8\%.}}
\fix{CCPNE identification times are included in the totals and shown in~\Cref{fig:mc_totaltime}.}
\texttt{Full-CCCE} is evaluated only for problem sizes for which the full $2^{nr}$-column formulation can be practically constructed and solved.

Coordination performance is evaluated using
\begin{equation}
    \textstyle C_{\rm avg}(z)
    :=
    \frac{1}{n} \sum_{i\in\mathcal N}C_i(z)
    \label{eq:additive_cost}
\end{equation}
and the Gini index used in \cite{rrce},
\begin{equation}
    \textstyle G(z)
    :=
    \frac{1}{2C_{\rm avg}(z)n^2}
    \sum_{i\in\mathcal N}\sum_{j\in\mathcal N}
    |C_i(z)-C_j(z)|.
    \label{eq:gini_index}
\end{equation}
A smaller Gini index indicates a more equitable allocation of cost across agents.

\subsection{Results}
\label{sec:results}

\subsubsection{Computational scalability}
\begin{figure}[hbt!]
    \centering
    \includegraphics[width=0.9\linewidth, trim = 0 1cm 0 0]{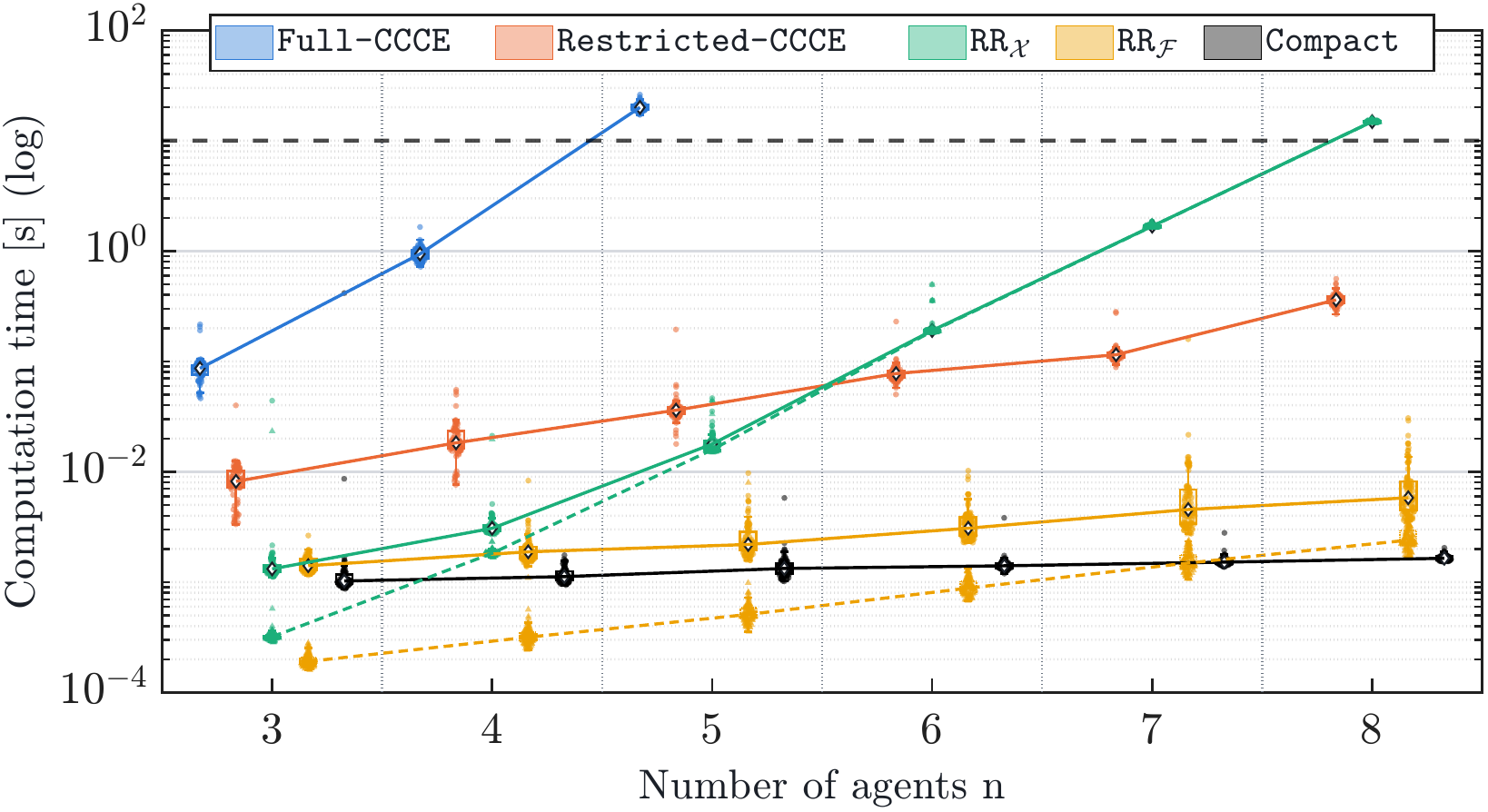}
    \caption{Computation time across Monte Carlo instances as the number of vertiports increases. Solid curves denote total computation time, while dashed curves denote CCPNE search time for \texttt{RR}$_{\mathcal X}$ and \texttt{RR}$_{\mathcal F}$. \fix{The black curve denotes \texttt{Compact} in~\Cref{rem:expected_allocation}.} The horizontal dashed line marks the $10$-second threshold used to truncate evaluation at larger problem sizes. Medians marked by $\diamond$.}
    \label{fig:mc_totaltime}
\end{figure}

The unrestricted formulations exhibit substantially steeper growth in computation time with $n$ than the restricted formulations, as shown in~\Cref{fig:mc_totaltime}. This difference reflects the reduction from the $2^{nr}$ unrestricted joint action space to the $n^r$ exact-one set. At $n=5$, the largest problem size evaluated for \texttt{Full-CCCE}, \texttt{Restricted-CCCE} reduces the median computation time from $20.0$~s to $0.038$~s, corresponding to a $99.81\%$ reduction and approximately a $524\times$ speedup. Once a formulation exceeds $10$s, larger problem sizes are not evaluated.

For \texttt{RR}$_{\mathcal X}$, the CCPNE search-time curve nearly coincides with the total computation-time curve, indicating that exhaustive equilibrium screening accounts for most of the computational cost. Restricting the candidate space from $2^{nr}$ to $n^r$ addresses this bottleneck. At $n=8$, the median CCPNE search time decreases from $14.9$~s for \texttt{RR}$_{\mathcal X}$ to $0.0018$~s for \texttt{RR}$_{\mathcal F}$, a $99.99\%$ reduction and approximately $8.4\times10^3$ speedup. 

\fix{The comparison shows that the reduced-rank representation reduces computation times, while the restriction yields more gradual growth with $n$. Combining both, \texttt{RR}$_{\mathcal F}$ achieves the lowest median total computation times among the four original formulations, by reducing the candidate-search space and optimizing over CCPNE mixture weights without explicit incentive constraints.}
\fix{For reference, \texttt{Compact} additionally exploits the expected-cost structure in~\Cref{rem:expected_allocation} and achieves the lowest median computation times overall.}

\begin{figure}[t!]
    \centering
    \vspace{2mm}
    \includegraphics[width=\linewidth, trim = 0 1.9cm 0 0]{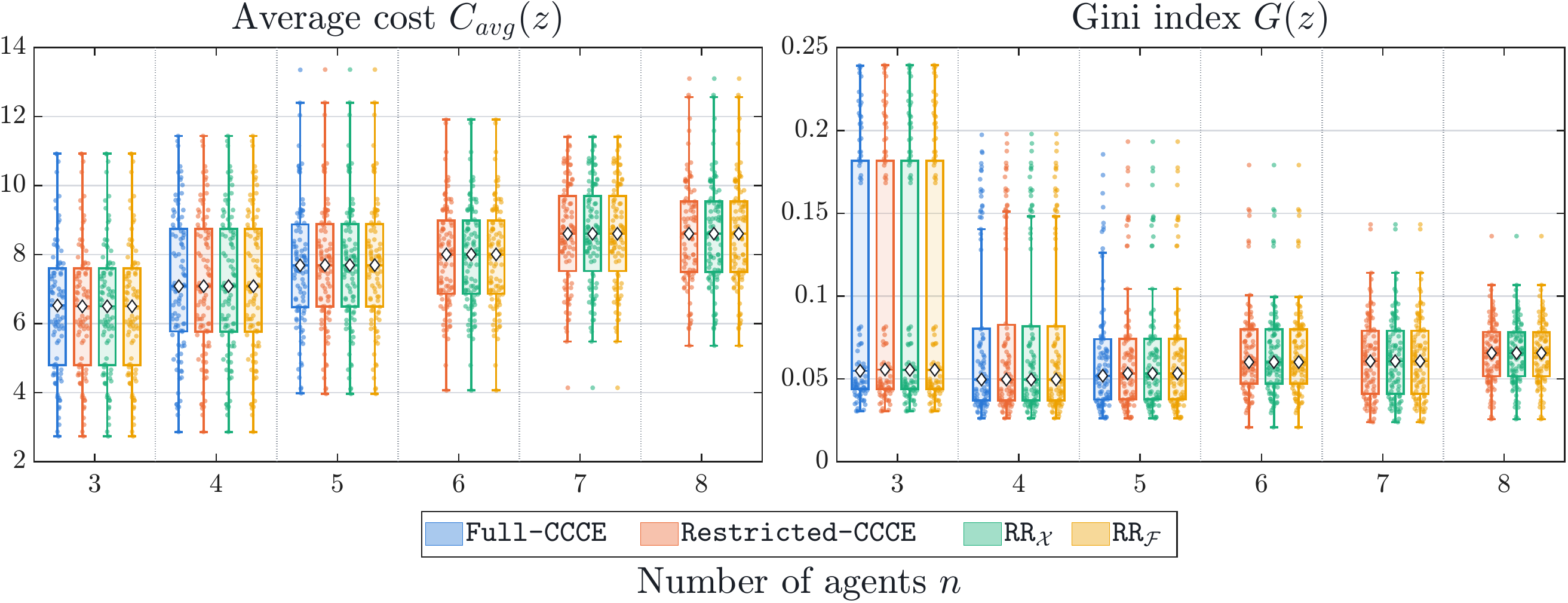}
    \caption{Coordination performance across Monte Carlo instances as the number of vertiports increases. The left panel shows $C_{\rm avg}(z)$, and the right panel shows the Gini index. Medians marked by $\diamond$.}
    \label{fig:mc_perf}
\end{figure}

\subsubsection{Coordination performance}

The computational gains are achieved without an observed loss of coordination performance, as shown in~\Cref{fig:mc_perf}. \texttt{Restricted-CCCE}, \texttt{RR}$_{\mathcal X}$, and \texttt{RR}$_{\mathcal F}$ attain identical average costs and Gini indices across the Monte Carlo instances. This agreement is consistent with the feasible-set equivalence established by \Cref{thm:restricted_geometry} and \Cref{prop:pne_completeness} under the CCPNE recovery condition \Cref{eq:pne_recovery_condition}. For the problem sizes where \texttt{Full-CCCE} is evaluated, it also yields identical cost and fairness values. 
\fix{This agreement with the unrestricted formulation is an empirical observation rather than a theoretical guarantee.}


\section{Conclusion}

We developed an exact-one resource-assignment restriction for \fix{CCCE} coordination in exclusive resource-assignment games. 
The restriction reduces the number of joint action probability variables from $\mathcal O(2^{nr})$ to $\mathcal O(n^r)$ and avoids prior CCPNE enumeration. 
We characterized its feasibility and derived a sufficient condition for preserving all CCPNEs in the unrestricted joint action space. 
Numerical experiments showed substantial computational savings without an observed loss in cost or fairness.

The restriction does not generally recover the unrestricted CCCE feasible set, \fix{and theoretical bounds on the potential loss from the reduced-rank representation remain open.} Future work will consider \fix{such bounds,} adaptive expansion of the restricted action set and more general assignment models that allow unused resources or resource-specific constraints.

\medskip

\bibliographystyle{IEEEtran}
\bibliography{reference}

\end{document}